\documentclass[twocolumn,showpacs,preprintnumbers,pra,hyperref]{revtex4}
\usepackage{amsfonts}
\usepackage{amsmath}
\usepackage{amssymb}
\usepackage{graphicx}
\usepackage[T1]{fontenc}

\begin{document}

\title{Slowing light with a coupled optomechanical crystal array}
\author{Zhenglu Duan and Bixuan Fan}
\affiliation{Center for quantum science and engineering, Jiangxi
Normal University, Nanchang, 330022, China } \affiliation{Center for
Engineered Quantum System, School of Mathematics and Physics,
University of Queensland, St Lucia, 4072, Qld, Australia}

\begin{abstract}
We study the propagation of light in a resonator optical waveguide
consisting of evanescently coupled optomechanical crystal array. In
the strong driving limit, the Hamiltonian of system can be
linearized and diagonalized. In this case we obtain the polaritons,
which is formed by the interaction of photons and the collective
excitation of mechanical resonators. By analyzing the dispersion
relations of polaritons, we find that the band structure can be
controlled by changing the related parameters. It has been suggested
an engineerable band structure can be used to slow and stop light
pulses.
\end{abstract}

\maketitle

\section{Introduction}

Owing to its important application in many fields, such as, low-threshold
lasing \cite{lasing}, pulse delay\cite{delay} and optical memories \cite%
{memory01,memory02,memory03}, slow light attracts a great deal of
practical interest. A number of schemes to delay and store light
have been suggested, such as electromagnetically induced
transparency (EIT) in the atomic
ensembles \cite{EIT01,EIT02}, photonic crystal waveguide band edges\cite%
{PC01,PC02}, solid-state multilayer semiconductor structure\cite{solid},
coupled resonator optical waveguide (CROW)\cite{CROW,Fan01,Fan02,FPcavity},
more complicated hybrid structure, e.g., coupled resonator optical waveguide
doped with atoms\cite{Sun01,Sun02,atom01,atom02}.

For a static photonic structure, for example a bare CROW, due to the
limitation of delay-bandwidth product constraint, it is not suitable to stop
light. To dynamically stop and release the light, a dynamically tunable
system is required. Fan suggested that, if there are extra resonators side
coupling to the optical cavity cells of the CROW, Fano interference can lead
to a large change of bandwidth of the system when a small refractive index
modulation is employed\cite{Fan01}. The velocity of light can therefore be
dynamically slowed down and even stopped. Unlike the case of EIT the light
is coherently stored in a static way in the resonance cavity array. Based on
this idea researchers have replaced optical resonators with atoms to couple
to the resonators in the CROW and found that the light can be converted to
collective excitations of atoms and then reversely converted and released%
\cite{Sun01,atom02}.

Optomechanics opens a door to directly control the mechanical motion with
light \cite{Review}. Many applications of optomechanics have been proposed,
for example, using cooled nanomechanical oscillators to test quantum
mechanics \cite{QM}, ultra-sensitive detection of force and dispalcement\cite%
{force,position}, quantum optics and quantum information processing \cite%
{squeezing,entanglement,statepreparation}.

Meanwhile, as a new quantum system, optomechanics is also used to
stop light. EIT effect in cavity optomechanical system with a
Bose-Einstein condensate (BEC) is suggested to slow the
light\cite{zhu}. Research groups led respectively by
Painter\cite{EITQM02} and Kippenberg\cite{EITQM01} proposed slowing
down light based on EIT in optomechanics. The photons are mapped
onto the phonon modes instead of internal atomic degrees of freedom
in the case of EIT in atomic ensembles. In fact, like the case of a
CROW, an optical waveguide coupled to an optomechanical crystal
array has been suggested to slow and stop light
pulse\cite{Painter1}.

Motivated by the work mentioned above, this paper investigates the photon
transmission in a homogeneous side coupled optomechanical crystal array, in
which each optical resonator in the bare CROW couples to an extra mechanical
resonator. The interaction between the mechanical mode and optical modes
allows the photonic band structure of CROW to be modulated, allowing
stopping and releasing light possible in our model. By adjusting the
refractive index of the photonic crystal, the photons can be mapped onto the
collective excitation of mechanical modes and then be stopped. Our scheme
offers a patternable, compact and on-chip platform to slow and stop light.

The paper is organized as follows: In Sec. II we present our model
of spatial periodic optomechanical crystal arrays, which consists of
optomechanical crystal cells side-coupled each other. After
linearizing the Hamiltonian, in Sec. III, we transform the system
into momentum space and then use Bogoliubov transformation to
diagonalise the Hamiltonian. Based on the dispersion relations of
upper and lower branch polaritons obtained in previous section, take
lower polariton for example, we investigate the band structure and
demonstrate how slow down the velocity of polariton by compressing
the bandwidth in Sec. IV. The final section concludes the paper.

\section{Derivation of model}

\begin{figure}[h]
\includegraphics[width=3.5in]{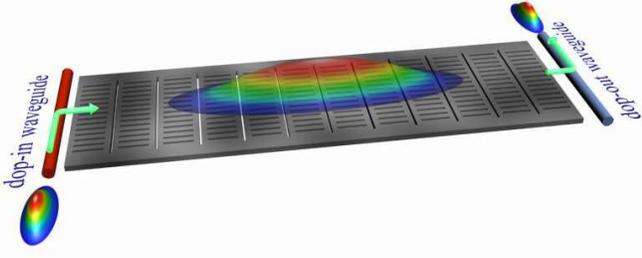}
\caption{{\protect\footnotesize The schematic of 1-D periodic array
of optomechanical crystals. Each optomechanical crystal cell is made
up of
photonic crystal and phononic crystal defect cavities coplanar\protect\cite%
{Painter}. The optomechanical crystal cells interact with each other by
evanescent light field between them. The light pulse drops in and out by
side-coupling wave guides.}}
\label{figure1}
\end{figure}
We consider an 1-D periodic array of optomechanical crystals, which
consists of $N$ evanescently coupled optomechanical crystals, shown
in Figure 1. The single optomechanical crystal cell, co-localizing
photonic and phononic resonances in a quasi one-dimensional
optomechanical crystal, is proposed by the research group led by
Painter\cite{Painter}. The mechanical modes of the optomechanical
crystal cell can be divided into common and differential modes of
in-plane and out-plane motion of these nanobeams. For simplicity, we
just consider the case that the gaps between the nanobeams are
time-independent, i.e. the common mode case. Therefore the coupling
between neighboring optical cavities is constant. To excite the
system, a probe optical signal is dropped in the optomechanical
array in a side-coupled configuration, and the output signal is
dropped out in a similar manner. With this consideration, the
Hamiltonian of the system in the reference frame rotating with probe
laser frequency $\omega _{p}$ can be written as

\begin{subequations}
\label{H}
\begin{align}
H& =H_{0}+H_{int}  \label{H1} \\
H_{0}& =\sum_{i}\hbar \delta \hat{a}_{j}^{\dagger }\hat{a}_{j}+\sum_{j}\hbar
\omega _{m}\hat{b}_{j}^{\dagger }\hat{b}_{j}  \label{H2} \\
H_{int}& =\sum_{j}\hbar g\left( \hat{b}_{j}^{\dagger }+\hat{b}_{j}\right)
\hat{a}_{j}^{\dagger }\hat{a}_{j}  \notag \\
& -\sum_{j}\hbar G\left( \hat{a}_{j}^{\dagger }\hat{a}_{j+1}+\hat{a}%
_{j+1}^{\dagger }\hat{a}_{j}\right) .  \label{H3}
\end{align}%
here $\hat{a}_{j}^{\dagger }\left( \hat{a}_{j}\right) $ and $\hat{b}%
_{j}^{\dagger }\left( \hat{b}_{j}\right) $ are creation
(annihilation) operators of optical cavity mode and mechanical mode
in the $j$-th optomechanical cell, respectively. $\delta =\omega
_{c}-\omega _{p}$ is the detuning between cavity field and probe
laser, $\omega _{m}$ is the mechanical resonator angular frequency,
the constant $g$ is the coupling strength between cavity and
mechanical resonator and $G$ denotes the nearest neighboring
evanescently coupling of intercavity.

When the intracavity fields have a large amplitude, i.e. in the
strong-driving limit, we can linearize the Hamiltonian by setting $\hat{f}%
=\left\langle f\right\rangle +\delta \hat{f}$ ($f=\hat{a}_{j},\hat{a}%
_{j}^{\dagger },\hat{b}_{j},\hat{b}_{j}^{\dagger }$), where $\left\langle
f\right\rangle $ is the steady mean value and $\delta \hat{f}$ is the
corresponding fluctuation around its steady value. With this ansatz, we then
obtain the linearized Hamiltonian

\end{subequations}
\begin{align}
H_{0}& =\sum_{j}\hbar \omega _{m}\delta b_{j}^{\dagger }\delta
b_{j}+\sum_{i}\hbar \tilde{\delta}\delta a_{j}^{\dagger }\delta
a_{j}\allowbreak  \label{H0} \\
H_{int}& =\sum_{j}\hbar \tilde{g}\left( \allowbreak \delta a_{j}^{\dagger
}\allowbreak \delta b_{j}+\delta a_{j}\delta b_{j}^{\dagger }\allowbreak
\right)  \notag \\
& -\sum_{j}\hbar G\left( \delta \hat{a}_{j}^{\dagger }\delta \hat{a}%
_{i+1}+\delta \hat{a}_{j+1}^{\dagger }\delta \hat{a}_{j}\right)  \label{HI}
\end{align}%
where the effective detuning is $\tilde{\delta}=\delta +g\left(
\left\langle b\right\rangle +\left\langle b^{\ast }\right\rangle
\right) /2$ and
the effective coupling between light and mechanical vibration is $\tilde{g}%
=g\left\langle a\right\rangle $. In Eq. (\ref{HI}) we have omitted the
counter-rotational wave term in the interaction between the cavity field and
mechanical vibration.

\section{Dispersion of polariton}

Let us study the Hamiltonian in $k$-representation. Taking into account the
periodic properties of the system, we can make Fourier transformations
\begin{subequations}
\label{AB}
\begin{align}
A_{k}& =\frac{1}{\sqrt{N}}\sum_{j}\delta \hat{a}_{j}e^{ikjL} \\
A_{k}^{\dagger }& =\frac{1}{\sqrt{N}}\sum_{j}\delta \hat{a}_{j}^{\dagger
}e^{-ikjL} \\
B_{k}& =\frac{1}{\sqrt{N}}\sum_{j}\delta \hat{b}_{j}e^{ikjL} \\
B_{k}^{\dagger }& =\frac{1}{\sqrt{N}}\sum_{j}\delta \hat{b}_{j}^{\dagger
}e^{-ikjL}
\end{align}%
where $A_{k}\left( A_{k}^{\dagger }\right) $ are the normal mode operators
of the coupled optical cavity, $B_{k}\left( B_{k}^{\dagger }\right) $ are
the boson operators to describe the collective excitation (phonon) of
mechanical resonators, $k=2\pi n/LN$ with $n=0,1...N-1$, and $L$ is the
distance of inter-cavity. Inserting above transformation relation into Eqs. (%
\ref{H1}), we arrive at the new Hamiltonian
\end{subequations}
\begin{align}
H& =\sum_{k}\hbar \omega _{m}B_{k}^{\dagger }B_{k}+\sum_{k}\hbar \omega
_{ph}\left( k\right) A_{k}^{\dagger }A_{k}  \notag \\
& +\sum_{k}\hbar \tilde{g}\allowbreak \left( \allowbreak A_{k}^{\dagger
}\allowbreak B_{k}+B_{k}^{\dagger }\allowbreak A_{k}\allowbreak \right)
\label{HK}
\end{align}%
here $\omega _{ph}\left( k\right) =\tilde{\delta}-2G\cos \left( kL\right) $
is the original dispersion property of photon dependent on quasimomentum $k$
in the side-coupled photonic crystal cavity array. Hamiltonian (\ref{HK})
describes the interaction of the photonic and phononic modes.

To decouple the Hamiltonian (\ref{HK}), we introduce the Bogoliubov
transformation

\begin{subequations}
\label{AKBK}
\begin{align}
A_{k}& =uC_{k}+vD_{k} \\
A_{k}^{\dagger }& =uC_{k}^{\dagger }+vD_{k}^{\dagger } \\
B_{k}& =vC_{k}-uD_{k} \\
B_{k}^{\dagger }& =vC_{k}^{\dagger }-uD_{k}^{\dagger }.
\end{align}
and the inverse transformation is
\end{subequations}
\begin{subequations}
\label{CKDK}
\begin{align}
C_{k}& =uA_{k}+vB_{k} \\
C_{k}^{\dagger }& =uA_{k}^{\dagger }+vB_{k}^{\dagger } \\
D_{k}& =vA_{k}-uB_{k} \\
D_{k}^{\dagger }& =vA_{k}^{\dagger }-uB_{k}^{\dagger }
\end{align}
Since operators $C_{k}$ and $D_{k}$ must satisfy Bosonic commutation
relations
\end{subequations}
\begin{align}
\left[ C_{k},C_{k}^{\dagger }\right] & =1 \\
\left[ D_{k},D_{k}^{\dagger }\right] & =1,
\end{align}%
transformation coefficients $u$ and $v$ have the relation $u^{2}+v^{2}=1$.
Substituting Eqs. (\ref{AKBK}) into Eq. (\ref{HK}), the Hamiltonian can be
rewritten as%
\begin{align}
H& =\left( \omega _{ph}\left( k\right) v^{2}+\omega _{m}u^{2}-2\tilde{g}%
uv\right) D_{k}D_{k}^{\dagger }\allowbreak  \notag \\
& +\left( \omega _{ph}\left( k\right) u^{2}+\omega _{m}v^{2}+2\tilde{g}%
uv\right) C_{k}C_{k}^{\dagger }  \notag \\
& +\left( \left( \omega _{ph}\left( k\right) -\omega _{m}\right) uv+\tilde{g}%
\left( v^{2}-u^{2}\right) \right) D_{k}C_{k}^{\dagger }  \notag \\
& +\left( \left( \omega _{ph}\left( k\right) -\omega _{m}\right) uv+\tilde{g}%
\left( v^{2}-u^{2}\right) \right) D_{k}^{\dagger }C_{k}.
\end{align}%
Obviously, if
\begin{equation}
\left( \omega _{ph}\left( k\right) \allowbreak -\omega _{m}\right) uv+\tilde{%
g}\left( v^{2}\allowbreak -u^{2}\right) =0
\end{equation}%
the Hamiltonian will have the diagonalized form

\begin{equation}
H=\sum_{k}\hbar \omega _{D}\left( k\right) D_{k}^{\dagger
}D_{k}+\sum_{k}\hbar \omega _{C}\left( k\right) C_{k}^{\dagger }C_{k}.
\label{Hk}
\end{equation}%
From Eqs. (\ref{CKDK}) and (\ref{Hk}) one can note that $C_{k}$ and
$D_{k}$ operators represent two types of elementary excitations
(phonon-photon polaritons) in coupled optomechanical array system,
which is the result of the coherently mixing of photons and phonons
through coupling in each optomechanics cell. The dispersion
relations of lower and upper branches polaritons are determined by

\begin{align}
\omega _{C}\left( k\right) & =\frac{1}{2}\left( \omega _{m}+\omega
_{ph}\left( k\right) -\sqrt{4\tilde{g}^{2}+\left( \omega _{ph}\left(
k\right) -\omega _{m}\right) ^{2}}\right) \\
\omega _{D}\left( k\right) & =\frac{1}{2}\left( \omega _{m}+\omega
_{ph}\left( k\right) +\sqrt{4\tilde{g}^{2}+\left( \omega _{ph}\left(
k\right) -\omega _{m}\right) ^{2}}\right) .
\end{align}%
It is found that the original single optical band structure is split
into two bands owing to the interaction between photons and phonons.

\section{Slowing light with tunable band structure}

The bandwidths of lower and upper branches, respectively, are
\begin{align}
\Delta _{WC}& =\omega _{C}\left( \pi \right) -\omega _{C}\left( 0\right)
\notag \\
& =2G+\Delta \\
\Delta _{WD}& =\omega _{D}\left( \pi \right) -\omega _{D}\left( 0\right)
\notag \\
& =2G-\Delta
\end{align}%
where

\begin{equation}
\Delta =\sqrt{\tilde{g}^{2}+\left( \Delta _{OM}/2-G\right) ^{2}}-\sqrt{%
\tilde{g}^{2}+\left( \Delta _{OM}/2+G\right) ^{2}}
\end{equation}%
with the detuning $\Delta _{OM}=\tilde{\delta}-\omega _{m}$.
Compared to the original optical band, the bandwidth of the lower
branch polariton is enlarged and the upper one is compressed.
Moreover, the most important thing is that both of these bandwidths
of lower and upper bands can be modulated by
changing parameters, such as $\tilde{g}$, $G$ and $\Delta _{OM}$. Figure \ref%
{figure2} shows a typical picture of the bandwidth of lower branch polariton
dependent on $\Delta _{OM}$, from which one can note that the bandwidth
decreases with increasing $\Delta _{OM}$. In more detail, when $\Delta
_{OM}/2\ll -\tilde{g}$, the bandwidth of lower band $\Delta _{WC}\simeq 4G$,
corresponding to the maximum bandwidth; when $\Delta _{OM}/2\gg \tilde{g}$,
the bandwidth is approximately equal to zero.
\begin{figure}[h]
\includegraphics[width=3.2in]{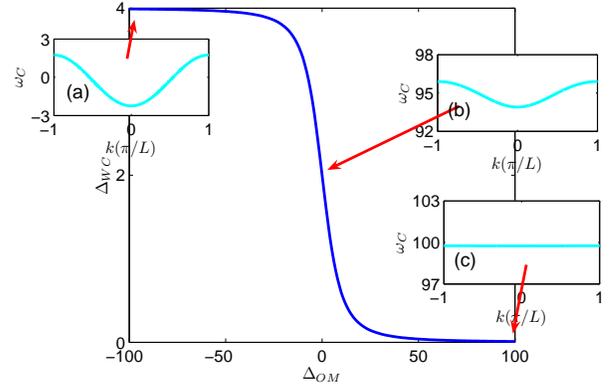}
\caption{{\protect\footnotesize The bandwidth of lower branch
polariton as a function of $\Delta _{0}$. The insets are the
corresponding band structures of polariton when detuning (a)$\Delta
_{0}=-100$; (b) $\Delta _{0}=0$ and (c) $\Delta _{0}=100$. $G=1$,
$g=5$, $\protect\omega _{m}=100$. Here $\Delta _{OM}$ is in units of
$G$.} } \label{figure2}
\end{figure}

\begin{figure}[h]
\includegraphics[width=3.2in]{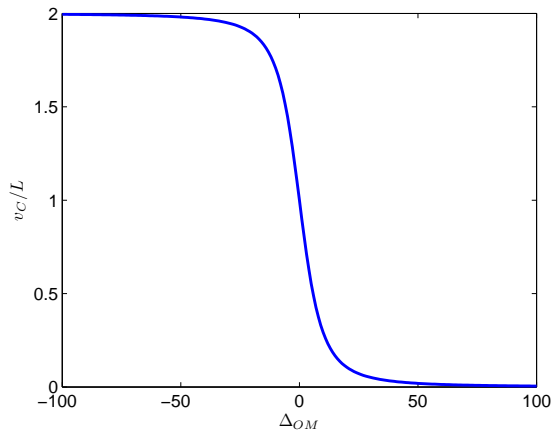}
\caption{{\protect\footnotesize Velocity of lower branch polariton.
Other parameters are the same as in Fig. \protect\ref{figure2}.}}
\label{figure3}
\end{figure}

On the other hand, it is known that the group velocity of polaritons in a
lattice is related with dispersion

\begin{eqnarray}
v_{C,D} &=&\frac{\partial \omega _{C,D}\left( k\right) }{\partial k}  \notag
\\
&=&-GL\sin \left( kL\right) \left( 1\pm \frac{\Delta _{OM}-2G\cos \left(
kL\right) }{\sqrt{4\tilde{g}^{2}+\left( \Delta _{OM}-2G\cos \left( kL\right)
\right) ^{2}}}\right) .
\end{eqnarray}%
which is also dependent on parameters $\tilde{g}$, $G$ and $\Delta
_{OM}$ and therefore can be tuned. Such a tunable band structure
leads to a tunable group velocity. Figure \ref{figure3} shows the
lower branch polariton with a momentum $kL=\pi /2$. We observe its
group velocity decreases rapidly from its maximum value $G$ and
vanishes as $\Delta _{OM}$ increases. In fact, such tunable band
structure can play an important role in optical communication and
quantum memory, for example, Fan suggested using a tunable CROW to
slow and stop light pulse\cite{slowinglight}.

\begin{figure}[h]
\includegraphics[width=3.2in]{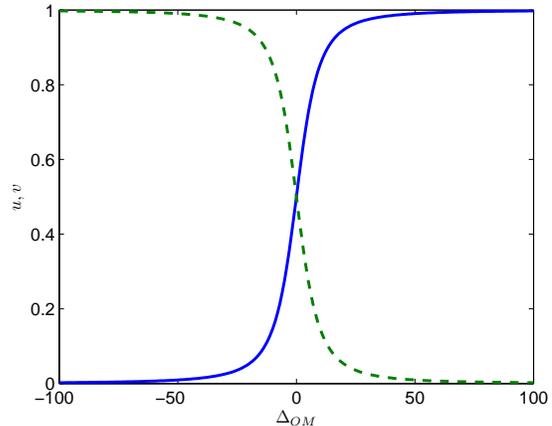}
\caption{{\protect\footnotesize Transformation coefficients in the
lower branch polariton versus detuning. The amplitude of $u$ and $v$
represent the ratio of photons and phonons in the lower branch
polariton. Solid and dashed lines represents $v $ and $u$
respectively. Other parameters are the same as in Fig.
\protect\ref{figure2}.}} \label{figure4}
\end{figure}

Here we briefly demonstrate the process, taking the lower branch
polariton as an example, to slow the light pulse in optomechanical
crystal array. To begin with, the optical cavity is adjusted to be
resonant with laser frequency, so the detuning is $\Delta
_{OM}/2=-\omega _{m}/2\ll -\tilde{g}$. At this point, the bandwidth
of lower branch is largest and can accommodate the entire light
pulse, and the lower branch polariton is made up of photons, shown
in figure \ref{figure4}. After the pulse enters completely into the
optomechanical array, we then compress the bandwidth of lower branch
polariton adiabatically by tuning the resonance frequency of the
optical cavity until $\Delta _{OM}/2=\omega _{m}/2\gg \tilde{g}$.
Further compressing the bandwidth, more and more photons are
converted to mechanical modes in the lower branch, meanwhile, the
velocity of polaritons slows down and approaches to zero.

From the point of conversion between photons and mechanical
collective excitations, one can also understand the mechanism of
stoping light. Because the total excitations number
$N_{k}=B_{k}^{\dagger }B_{k}+A_{k}^{\dagger }A_{k}$ commutes with
Hamiltonian $H$, when adjusting some parameters, such as $\Delta
_{OM}$, the total excitations number is conserved, while the numbers
of photons and mechanical collective excitations $A_{k}^{\dagger
}A_{k}$ and $B_{k}^{\dagger }B_{k}$ are not conserved due to not
commuting with the Hamiltonian. Hence the photons and mechanical
collective excitations are mutually convertible, which results in
mapping the light onto the mechanical vibration and vice versa.
Figure \ref{figure4} illustrates that the conversion between the
photons and mechanical collective excitations. The number of
mechanical collective excitation increases, from zero to unity,
while the number of photons decrease from unity to zero, with
increasing the detuning $\Delta _{OM}$.

\begin{figure}[h]
\includegraphics[width=3.2in]{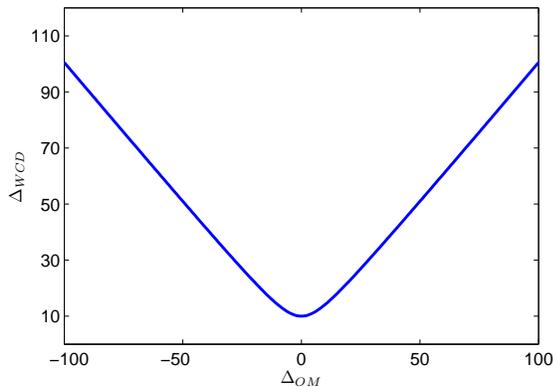}
\caption{{\protect\footnotesize Band gap between upper and lower
branches versus detuning. There is a minimum value when
$\Delta_{OM}=0$. Other parameters are the same as in Fig.
\protect\ref{figure2}.}} \label{figure5}
\end{figure}

We keep in mind that the rate of tuning cavity frequency should be less than
the band gap between upper and lower branches, which is given by

\begin{equation}
\Delta _{WCD}\left( k\right) =\omega _{D}\left( k\right) -\omega _{C}\left(
k\right)
\end{equation}%
This limitation avoids the polaritons in lower branch jumping up to the
upper branch. Figure \ref{figure5} shows that the band gap first linearly
decreases and then linearly rises with increasing the detuning $\Delta _{OM}$%
. The minimum value of band gap is about $2G$ when the detuning of optical
cavity is resonant with mechanical resonator.

To practically stop light in the coupled optomechanics array, we
tune the detuning between optical cavity field and probe laser by
adjusting the refractive index of material,e.g. Si, which makes up
of the optomechanical crystal. In our scheme, the amplitude of
detuning modulation is the same order of magnitude of the mechanical
resonance frequency, so the refractive index shift should be

\begin{equation}
\frac{\delta n}{n}\approx \frac{\delta \omega }{\omega _{0}}\sim 10^{-5}
\end{equation}%
which is feasible in practical optoelectronic devices\cite{modulation}. Here
we have taken the typical parameters $\omega _{c}/2\pi =200$ THz, $\omega
_{m}/2\pi =10$ GHz\cite{Painter}.

\section{Conclusion}

We have studied the model of light transmission in a spatial
periodic optomechanical crystal array. The optical cavities of the
array evanescently couple to each other one by one to form a CROW.
In the strong driving limit, we linearized the system and obtained
the dispersion relations of lower and upper branch polaritons with
Bogoliubov transformation in the momentum space. Our results show
that the modulation of detuning between optical cavity and laser
light can vary the bandwidths of polaritons, which has been
demonstrated to be able to stop and release a light pulse.

\section{Acknowledgments}
We thank Andrew Bolt for polishing English.

\end{document}